\documentclass{nature}
\usepackage{amsmath}
\usepackage{amssymb}
\usepackage{wasysym}
\usepackage[dvips]{graphicx}
\usepackage{cancel}
\usepackage{ulem}
\usepackage{xcolor}
\usepackage[latin1]{inputenc}
\usepackage{times}
\title{\textit{Well-defined sub-nanometer graphene ribbons synthesized inside carbon nanotubes}}

\author{Hans Kuzmany$^{2*}$, Lei Shi$^{1,2*}$ Sofie Cambré$^3$, Miles Martinati$^3$, Wim Wenseleers$^3$, Jen\H o K\"urti$^4$, J\'anos Koltai$^4$, Gerg\H o Kukuczka$^4$, Kecheng Cao$^5$, Ute Kaiser$^5$, Takeshi Saito$^{6}$,  \& Thomas Pichler$^{2}$}

\begin{document}

\maketitle

\vspace{0.5cm}

\begin{affiliations}

\item School of Materials Science and Engineering, State Key Laboratory of Optoelectronic Materials and Technologies, Nanotechnology Research Center, Sun Yat-sen University, Guangzhou 510275, Guangdong, P. R. China
\vspace{0.2cm}

\item Faculty of Physics, University of Vienna, 1090 Wien, Austria
\vspace{0.2cm}

\item Experimental Condensed Matter Physics Laboratory, Physics Department, University of Antwerp, B-2610 Antwerp, Belgium 
\vspace{0.2cm}

\item  Department of Biological Physics, ELTE E\"otv\"os Lor\'and University, P\'azm\'any P\'eter stny. 1/A, 1117 Budapest, Hungary
\vspace{0.2cm}

\item Central Facility for Electron Microscopy, Electron Microscopy Group of Materials Science, Ulm University, Ulm 89081, Germany
\vspace{0.2cm}

\item Nanotube Research Centre, National Institute of Advanced Industrial Science and Technology (AIST), 305-8565 Tsukuba, Japan 

\end{affiliations}

\vspace{0.5cm}

\clearpage
\begin{abstract}
Graphene nanoribbons with sub-nanometer widths are extremely interesting for nanoscale electronics and devices as they combine the unusual transport properties of graphene with the opening of a band gap due to quantum confinement in the lateral dimension. Strong research efforts are presently paid to grow such nanoribbons. Here we show the synthesis of 6- and 7-armchair graphene nanoribbons, with widths of 0.61 and 0.74 nm, and excitonic gaps of 1.83 and 2.18 eV, by high-temperature vacuum annealing of ferrocene molecules inside single-walled carbon nanotubes. The growth of the so-obtained graphene nanoribbons is evidenced from atomic resolution electron microscopy, while their well-defined structure is identified by a combination of an extensive wavelength-dependent Raman scattering characterization and quantum-chemical calculations. These findings enable a facile and scalable approach leading to the controlled growth and detailed analysis of well-defined sub-nanometer graphene nanoribbons.
\end{abstract}

\vspace{1.5cm}
\section{Keywords}
graphene nanoribbons, electronic structure, Raman scattering, resonance profiles, Albrecht theory, GW calculation

\maketitle

\clearpage 

\section{Introduction}
Sub-nanometer graphene ribbons (graphene nanoribbons, GNRs) are promising structures for future 
electronic devices\cite{Bennett13APL,Martini19C} as they are considered of unifying the 
unique electronic properties of graphene \cite{Novoselov04s} with a reasonably sized gap in their electronic structure. The gap results from
quantum confinement in the lateral direction and consequently scales with the inverse width of the ribbons. 
\cite{Son06PRL,Barone06NL} The ribbons are strips of 
carbon atoms cut out from a graphene lattice. At present most common strips are of the armchair type, i.e., 
the edge of the ribbons consists of coaxial carbon pairs oriented parallel to the direction of the ribbon axis 
(armchair graphene nanoribbons (AGNR)). They are characterized by the number n of such pairs across their width. The electronic structures of AGNR can be classified into n= 3p, 3p+1 and 3p+2 species, where p is an 
integer.\cite{Son06PRL} 
%At present such nanoribbons have been prepared with a reasonably large number of widths 
%\cite{DiGiovannantonio18JACS}.  
%(STS) has been used frequently to determine the gap in the electronic spectrum (band gap)\cite{Merino-Diez17AN,Chen13AN}.
 One of the most commonly investigated structures is the n=7 AGNR (7-AGNR) \cite{Cai10N,Talirz16AM} with 
band gaps between 2.1 and 2.3 eV as reported from scanning tunneling spectroscopy (STS) and optical studies. 
\cite{Denk14NC}
More recently ribbons with more complex topologies were 
grown which have coved zigzag, chevron or chiral type structures.\cite{Talirz16AM,Durr18JACS,Sun20AM}
The ribbons are usually grown 
on Au substrates from preselected and properly designed flat poly-aromatic hydrocarbon molecules 
(PAHs).\cite{Gigli19AN} In this case the PAHs are vacuum deposited on the substrate and subsequently  transformed to 
polymeric units with nanoribbon structure. To take advantage of the ribbons grown in this way subsequent transformation to insulating substrates is necessary. \cite{Overbeck19PSSB} Raman scattering was used to evidence that the nanoribbons do not suffer in quality by this transformation.
%The structure of the edges along the ribbons determines the electronic character of the ribbons \cite{Groning18N,Pedramrazi18NL}.
%Recently GNR are prepared from bottom up chemical reactions using flat poly-aromatic 
%hydrocarbon molecules (PAHs) as precursor materials \cite{Merino-Diez17AN,Narita19CS}. 
%11sCheng,17sZhang
\par
 In general the carbons at the edge of the ribbons are saturated by hydrogen but by selecting special PAHs, bandgap engineering \cite{Verbitskiy16NL,Pedramrazi18NL,Narita19CS}, construction of ribbon heterojunctions
\cite{Chen15NN,Cao17PRL} and ribbons with unusual electronic properties\cite{Groning18N} have been demonstrated where topological properties of the ribbons under investigation play an important role.  Most recently ribbons became relevant for applications in photocatalytic hydrogen generation.\cite{Akilimali20CT}
\par
Besides STS and electron microscopy, Raman scattering has repeatedly been used to characterize GNR. Several Raman active vibrational modes 
were identified to characterize the ribbons. Such modes are among others the radial breathing like mode (RBLM), the CH in plane bending mode (CH-ipb), the D line, and the graphene G line.\cite{Overbeck19PSSB,Verbitskiy16NL} 
The RBLM frequency scales with (the square root of) the inverse ribbon width,\cite{Vandescuren08PRB,Liu20Carbon} and is the pendant to the 
radial breathing mode of the carbon nanotubes.
%In fact RBLM frequencies have been derived 
%theoretically from e. g. a zone folding process of graphene \cite{Gillen10PSSB}. 
%The D line can be considered as the zone edge vibration of graphene which 
%becomes observable in Raman experiments of ribbons from zone folding. 
\par
Filling and consecutive chemical reactions inside single-walled carbon nanotubes (SWCNTs) is a promising technique to grow new nanoscale materials in general.\cite{Lim13NC,Khlobystov11AN,Kuzmany17PSSRL}  
%\cite{Kharlamova16APAS&P}
In addition, the one-dimensional geometry of the CNTs is an excellent template for the controlled growth of conventional or exotic low-dimensional compounds.\cite{Kitaura09ACIE,Shi16NM,Pham18S}. Here, we show that this in-tube synthesis can be used to grow GNRs with well-defined sub-nanometer width. Growth of such narrow ribbons with controlled width has previously been limited to the polymerization of specific PAHs on Au substrates.\cite{Chamberlain12AN} Although providing well-defined freestanding ribbons with relatively high yield, the flexibility of preparing GNRs with different widths and structures is limited by the availability of appropriate precursor molecules. 
When synthesizing GNRs inside SWCNTs, the width of the GNRs is determined by the diameter of the SWCNTs, hence could in the future be flexibly tuned by starting from CNT samples with different diameters (eventually even diameter/chirality-sorted SWCNTs). Moreover, the hybrid structures of encapsulated ribbons and a semiconducting SWCNT, both with a different band gap can be interesting also from a materials design perspective.

\par
The growth of GNRs inside SWCNTs has been demonstrated by first filling the SWCNTs with flat precursor molecules, such as coronene\cite{Chernov13AN,Anoshkin14CPC} or other PAHs\cite{Chamberlain12AN}, and subsequently transforming them at elevated temperature. Although these initial results were very promising, growth of high-quality specific types of GNRs with well-defined widths still remained a challenge. In particular the analysis of the electronic properties of the objects inside the tubes remained difficult due to the overlapping electronic transitions of the carbon nanotubes themselves. To avoid this overlap, functionalization was necessary of the otherwise pristine tubes.\cite{Lim15AN} As it will be demonstrated below, Raman scattering, in particular in combination with evaluation of resonance Raman excitation profiles, is an excellent tool to reveal the electronic structure of the objects inside the tubes. This is due to the double selective nature of the resonance Raman scattering process. It is selective with respect to the geometrical structure of the objects by the vibrational mode but also selective with respect to the electronic structure by the resonance profile. Small width of the Raman lines indicates well-defined geometrical structures and sharp excitation profiles provide evidence for uniform electronic configuration of the GNRs.
%As it will be demonstrated below, wavelength-dependent Raman scattering is an excellent tool to characterize the encapsulated objects, since it is selective in two different complementary ways to their structure, providing simultaneous access to both the characteristic vibrational and electronic resonances of the nanoribbons, thereby also automatically disentangling the signals from the nanoribbons from those of the surrounding nanotubes.
\par
In this work, we demonstrate the synthesis of two specific GNRs, i.e., 6- and 7-AGNRs, with well-defined geometrical structure, from the bulky molecule ferrocene (FeCp$_2$). Evidence for the growth of the AGNRs inside the tubes comes from aberration-corrected high-resolution transmission electron microscopy (AC-HRTEM) and from Raman scattering. We also show that wavelength-dependent Raman spectroscopy gives direct access to the electronic and optical properties of the encapsulated objects by the evaluation of resonance Raman excitation profiles.  This allows determining experimentally the excitonic gap and even more the electronic structure beyond the gap. High level first principle calculations provide for the first-time relative intensities of the Raman lines of the GNRs by using the Placzek formulation with energy dependent polarizabilities.\cite{Kukucska19arX}  The latter procedure is a fundamental progress in the evaluation of Raman intensities.

\section{Experimental Results and Analysis}

\begin{figure*}[tbp]
\includegraphics[width=0.8\linewidth]{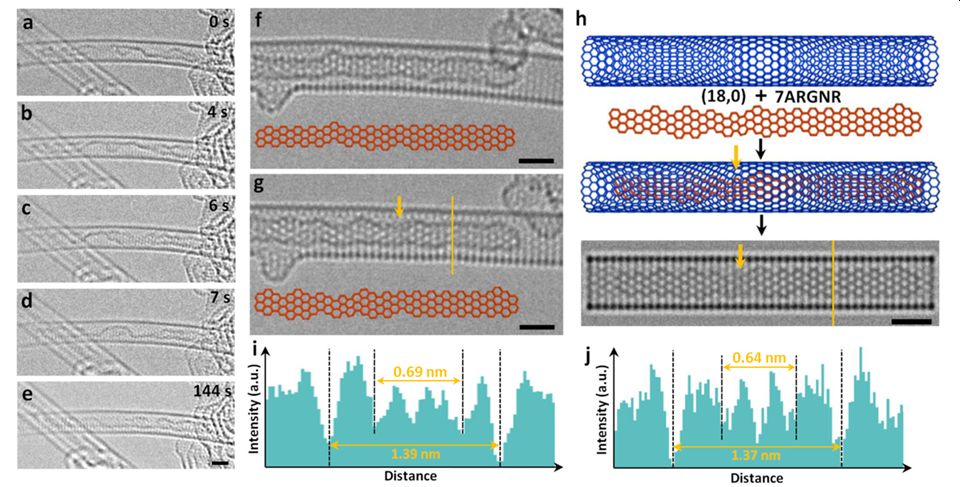}
\vspace{0.3cm}
\caption{AC-HRTEM images and corresponding simulations for AGNR@SWCNTs. a-e, Time series of images showing a ribbon confined in a SWCNT twisting under 80 keV electron beam. The modulation of the response with time is evidence for a flat ribbon. f,g, Two typical AC-HRTEM images in atomic resolution showing a flat 7-AGNR dominated structure confined in a SWNT. Scale bar is 1\;nm. The inserted models in red present the corresponding structure of the ribbon, showing also the instability of the ribbon under electron beam irradiation (i.e., f-g are the same ribbon but acquired after different exposure times). h, TEM image simulation for confirming the structure of 7-AGNR@SWCNT in g. It depicts the tube (18,0), the ribbon, the combination of the two and the simulation. The most part of 7-AGNR and the wall of SWNT are AA stacked showing clear graphene structure for the former, while a small mismatch part as indicated by yellow arrows results in blurred contrast in the simulated TEM and in the raw TEM image in g.  i,j, Intensity profiles along the yellow line of the recorded pattern i and of the simulated pattern j. The round (outer) tubes exhibit a strong reduction of contrast at the edge whereas the contrast is weaker at the edge of the flat 
ribbons. Indicated distances correlate with the diameter of a (18,0) tube and evidence a net ribbon width between 0.64 and 0.69\;nm.  
}
\label{fig:TEM}
\end{figure*}

{\bf Aberration-Corrected High Resolution Transmission Electron Microscopy (AC-HRTEM)} The standard feature for the observation of encapsulated ribbons in TEM is an alternating pattern of narrow 
and wide signals from the material inside.\cite{Chuvilin11NM,Chamberlain12AN} It originates from an electron beam induced twisting of parts of the ribbons.  
Figure\;\ref{fig:TEM} depicts a collection of results from FeCp$_2$ filled and subsequently transformed SWCNT species as acquired by the SALVE instrument with Cs/Cc aberration corrector. 
It demonstrates the twisting of the graphene ribbons by the electron beam (panel a-e) and provides images of the ribbons in atomic resolution with corresponding simulations (panel f-j). The time-dependent AC-HRTEM images show clearly that the observed species are flat objects, that twist under the influence of the electron beam, and cannot be identified as inner nanotubes. The AC-HRTEM images (panels f-j) provide clear evidence of the atomic structure of the ribbons, however, also demonstrate that the nanoribbons are very unstable under the electron beam and defects along the length of the structure are quickly generated by the beam. Therefore, these AC-HRTEM images are not representative for the quality of the as-grown ribbons and only serve to evidence that the structures are flat ribbons. Evidence for the well-defined electronic structure of the as-grown ribbons comes from our detailed wavelength-dependent Raman spectroscopic experiments (see below).  More details on the AC-HRTEM are provided in the Supporting Information Section (a).
%{\it The dominating species in this case are 7-AGNR. In detail e. g. subfigure f depicts extended 7-AGNR, despite some %defects which probably were induced by the imaging electron beam. The ribbons are interfaced to a structural and %topological different 6-AGNR. Subfigure g depicts an example where again extended species of 7-AGNR are observed %connected to more defective sections with even short segements of 8-AGNR.}  
%More details on the geometry, on the chirality of the involved tubes, and on the stacking of the ribbons are provided in the Supporting Information Section (a).

%

%%%%%%%%%%
%%%%%%%%%%
%%%%%%%%%% Temperature dependence, Raman

\par
%Raman scattering is another important technique to characterize carbon structures, because this non-invasive technique allows for investigating the GNRs 
%inside the SWCNT. When measuring the Raman response of different SWCNT samples, filled with ferrocene
% and heated in vacuum at different temperatures with 50C steps, the Raman response is observed to be very sensitive to the transformation temperature. 
\clearpage
{\bf Raman Scattering} Previously, we reported the observation of a set of Raman lines after thermal conversion of 
FeCp$_2$ filled SWCNT \cite{Kuzmany17PSSRL,Kuzmany15PSSB} but the origin of these lines remained unclear as they did not fit to proper model calculations
and high resolution TEM was not available. Here we identified the origin of the Raman lines which turned out to be very sensitive to transformation temperature.
Figure\;2a-d depicts this behavior in a plot where Raman intensities are characterized by a color code as a function of 
Raman frequency and transformation temperature. Raman spectra were normalized to the 2D band which is least influenced by the changes induced by the GNR encapsulation. 
Figure\;2a,b depicts the responses measured for 568\;nm excitation while Fig.\;2c,d presents those for 633\;nm excitation. 
At low temperature, only the RBMs (around 200\;cm$^{-1}$) and the G-line of the SWCNTs (around 1600\;cm$^{-1}$) are observed. When increasing the transformation 
temperature beyond 500\;$^\circ$C, depending on the excitation laser used
two groups of Raman lines can be observed. They exhibit a maximum Raman intensity between 600 and 700\;$^\circ$C, indicated by the horizontal white arrows in the figure. As shown further on, these two groups of lines correspond to the in-tube synthesis of AGNRs with different widths. 
%which turned out to be relevant for an unambiguous assignment of transformation temperatures.
%The maxumum transformation temperature turned out to depend on the energy of the exciting laser. The higher energy laser yields slightly lower values for optimum transition
%for both observed species. 
%For transformation at 700\;$^\circ$C and excitation with the red laser lines from the new objects are as high as the response from the tubes. For transition temperatures between 600\;$^\circ$C and 700\;$^\circ$C both groups of lines can be observed albeit preferably with lasers of different wavelength.

\begin{figure*}[tbp]
\includegraphics[width=\linewidth]{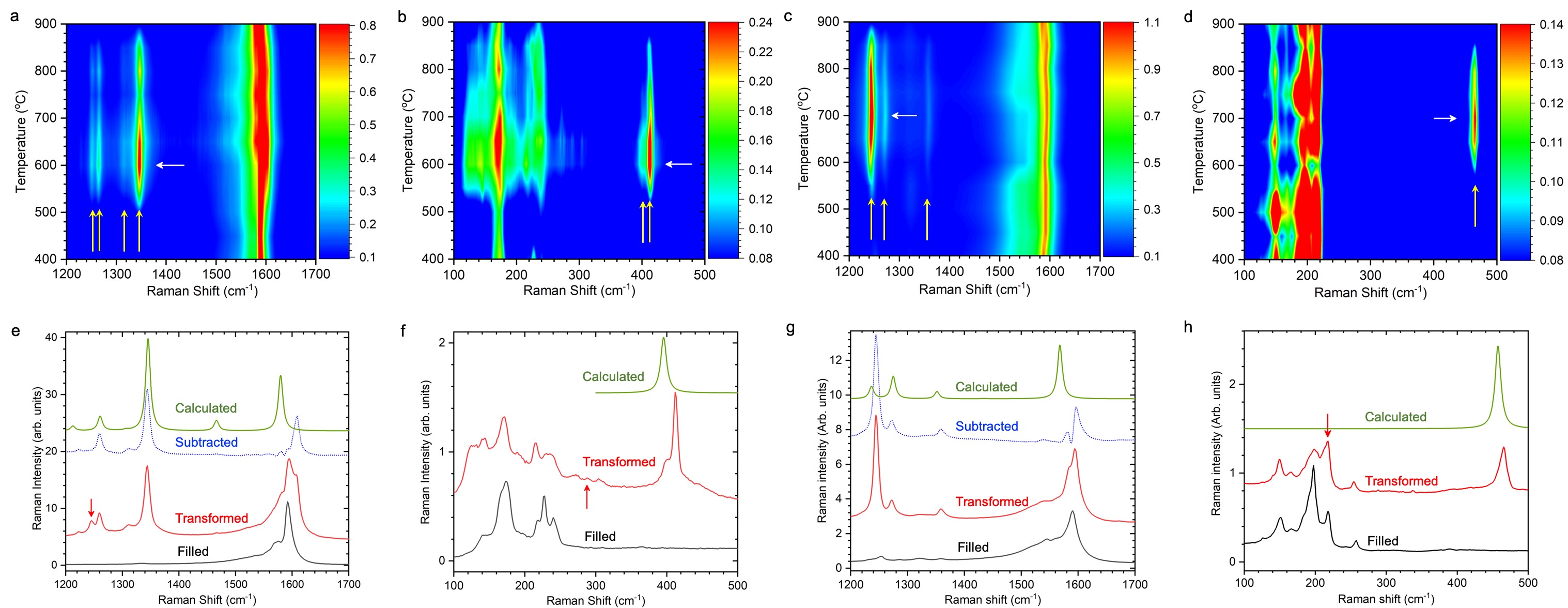}
\vspace{0.2cm}
\caption{ Raman scattering for temperature dependent transformation. a-d, Color-code maps for Raman
 line intensities observed after transformation of 
FeCp$_2$@SWCNTs as a function of Raman frequency and transformation temperature. The latter was increased in 50\;$^\circ$C steps. Vertical white arrows highlight the positions of the main new Raman 
lines. Horizontal white arrows are located at temperatures of maximum Raman response. Raman spectra in a and b 
and in c and d were excited 
with a yellow laser (568 nm) and red laser (633 nm), respectively. e-h, Raman spectra for FeCp$_2$@SWCNT tuned by transformation temperature to optimized 
response for the two groups of lines. e, Spectra as observed for yellow laser excitation at 568 nm in the high frequency region, 
from bottom to top: tubes filled with FeCp$_2$ (black) and subsequent transformation at 600 $\circ$C (red), AGNR contribution by subtracting the Raman signals from the nanotubes (blue, dotted), and calculation (green)
for 7-AGNR. The Raman line marked by a down-arrow was also subtracted since it originates form 6-AGNR as discussed below. f, Raman signals from 
low frequency region, from bottom to top: filled tubes, transformed at 600\;$^\circ$C, and calculated for 7-AGNR. 
The arrow marks the response from DWCNTs 
obtained during the transformation process. g,h, Similar spectra as in e,f but for transformation temperatures of 700\;$^\circ$C, 
recorded with 633\;nm laser, and as calculated for 6-AGNR. 
}
\label{fig:Temp}
\end{figure*}
 The lower panels of Fig.\;\ref{fig:Temp} depict the Raman spectra explicitly. Panels e and f are spectra recorded at 568\;nm 
(transformation temperature 600\;$^\circ$C)
and g and h are spectra recorded at 633\;nm (transformation temperature 700\;$^\circ$C), for two different 
Raman frequency regions. When comparing the Raman spectra of the ferrocene-filled and temperature-transformed 
samples (red) with the spectra of filled SWCNTs without transformation (black), the new Raman features appear
 very well separated from the response of the SWCNTs in general. In the high-frequency region, the 
 spectra of the SWCNTs can be subtracted straight forwardly, yielding the Raman response of the newly synthesized objects (blue dotted spectra in panels e and g). 
For the low-frequency region, a subtraction is more difficult, since during the transformation also inner tubes are formed thereby changing the response of the SWCNTs as well (red arrows). Even though, due to their  higher frequency, the lines from the RBLMs of the GNRs can be well separated from the response of the CNTs. 
The figures demonstrate very narrow line widths of the order of 10\;cm$^{-1}$ for the response of the nanoribbons. This linewidth is equivalent to the Raman response of high-quality ribbons grown on Au substrates and subsequently transferred to semiconducting substrates.\cite{Borin19ACSANM,Martini19C,Overbeck19PSSB} Such narrow line widths (in combination with a clear resonance profile) can only be observed for nanoribbons with a well-defined width along the entire length, as width variations would lead to inhomogeneous broadening of the lines or even disappearance of the RBLM mode. Figure\;2 also presents the calculated Raman spectra for 7-AGNR species (panels e and f, in green) and for 6-AGNR (panels g and h, in green),
 which were obtained from first principle calculations using the dynamical Placzek formalism \cite{Kukucska19arX}. More details about these calculations are in the Method section and in the 
 Supporting Information Sections (b). 
 They correspond very well to the observed experimental lines, even in 
relative intensity. A blown up version of this comparison can be found in the Supporting 
Information Section (d). For the 7-AGNR, the high-frequency modes are accordingly identified as the CH-ipb mode 
(1258\;cm$^{-1}$), the D-line (1344\;cm$^{-1}$) and the G line (1606\;cm$^{-1}$). The low frequency mode at 414\;cm$^{-1}$ (with small shoulder at 400\;cm$^{-1}$) can be identified as the RBLM. Likewise, for the 
6-AGNR the RBLM is located at 465\;cm$^{-1}$, the CH-ipb mode at 1243\;cm$^{-1}$, the D-line at 1358\;cm$^{-1}$ and the G line at 1595\;cm$^{-1}$. 
Note that the latter is strongly overlapping with the G line of the SWCNTs which prevents its direct determination 
from the difference (blue) spectra. However, the frequency can be obtained by measuring the 
sum of the CH-ipb and the G line of the ribbons which is observed at 2839\;cm$^{-1}$. This localizes the G line at 1595\;cm$^{-1}$.   

\begin{table}
  \caption{ Raman lines of 7-AGNR and 6-AGNR; Frequencies ($\omega_{ph}$), intensities (weak (w), medium (m), strong (s), very strong (vs)), and linewidths W (FWHM, in parentheses) 
	for the observed Raman lines (column Exp.) as 
compared to calculation (column Calc.) and to reference (column Ref.). For 7-AGNR and 6-AGNR the latter are 
from reference\cite{Cai10N} and reference\cite{Basagni15JACS}, respectively. The column with 
the calculated frequencies depicts also the difference to the experiment in \%. The last two columns depict the excited state 
frequencies ($\omega_{ph}^*$) and 
the Huang-Rhys factor (HR) for the first excited state, both obtained from a fit of the resonances to the Albrecht A-term. 
All frequencies are given in cm$^{-1}$ and rounded to integer values.
	}
	\tabcolsep0.5mm
  \begin{tabular}[htbp]{@{}llllll@{}}
    \hline
    Mode & $\omega_{ph}$ (W)& $\omega_{ph}$ &$\omega_{ph}$ (W)&$\omega_{ph}$$^*$&HR \\
				&Exp.&Calc.;\% & Ref.\cite{Cai10N,Basagni15JACS}&from fit&from fit\\
				&(cm$^{-1}$) &(cm$^{-1}$) &(cm$^{-1}$)&(cm$^{-1}$)&(cm$^{-1}$)\\
    \hline
		7-AGNR&&&&&\\
		\hline
    RBLM &414m (7.5) &395m; 4.5&395m (22.1)&434&0.08  \\
		%&1222w&1212w; 0.8&&&\\
		CH-ipb &1258s (9.6) &1261s; 0.2&1263s (30)&1195&0.25 \\
		%&1312w&1305 w; 0.5&&&\\
		D&1342vs (10.8)&1345vs; 0.2&1344vs (26)&1275&0.25\\
		%&1466w&1466 w; 0&&&\\
		G&1607s (13)&1580 s; 1.7&1607vs (31)&no fit&\\
    \hline
		6-AGNR&&&&&\\
		\hline
		 RBLM &466m (8) &457m; 1.7&not reported&&  \\
		CH-ipb &1243vs (9.6) &1236m; 0.6&1245s (100)&1281&0.17 \\
		      &1272m  (10.6)       &1272s; 0&&1272&0.25\\
		D     &1358m  (11.4)      &1352w; 0.4 &1315m&1358&0.25\\
		%      &1366w&1366w; 0&&&\\
		G     &1595 &1568s; 1.7&1590vs &no fit&\\   
		\hline
  \end{tabular}
  \label{Tab:1}
\end{table}

Table 1 presents the experimental Raman frequencies and line widths for both the 6-AGNR and 
7-AGNR and compares it to our theoretical values as well as to values for the same modes measured 
on Au-substrates for 7-AGNR.\cite{Cai10N} For the 6-AGNR a comparison is made with the only reported data originating from 
the fusion of linear chains of poly-paraphyenlye at 800\;K.\cite{Basagni15JACS} The agreement between calculated and 
experimentally determined Raman frequencies in the high-frequency region is better than 2\% and also the agreement between calculated and experimental relative line intensities is excellent, except for the CH-ipb of the 6-AGNR which appears too weak in the calculation. The table highlights the unusually narrow width of the Raman lines from the encapsulated species. In three very recent reports, Raman spectra for 7-AGNR grown on Au substrates and subsequently transferred to transparent substrates are shown to exhibit similar narrow line widths as 
observed here.\cite{Borin19ACSANM,Martini19C,Overbeck19PSSB}
%For transformations in the temperature range between 600\;$^\circ$C and 700\;$^\circ$C Raman lines can be observed for 7-AGNR and 6-AGNR.
%%%%%%%%%%%%%%%%%%%Resonance, Map & cross sections

\begin{figure*}[tbp]
\includegraphics[width=0.95\linewidth]{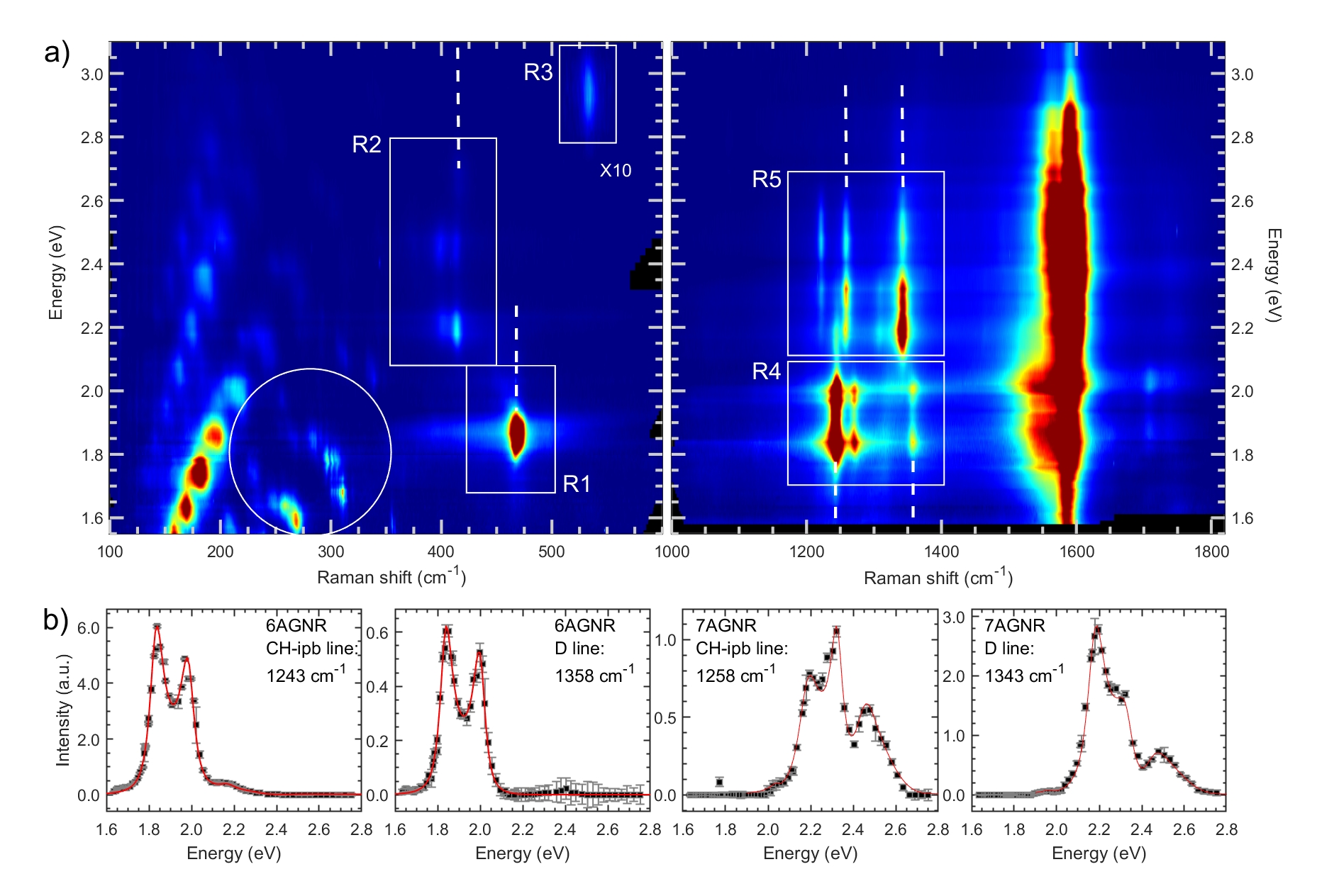}
\caption{Wavelength-dependent Raman spectroscopy of AGNR@SWCNT. a, Two-dimensional Raman map for the resonance excitation of AGNR@SWCNT. The features from the ribbons are framed into the white rectangles with numbers R1 to R5. The rectangles R2 
%(zoomed version shown in Figure 5) 
and R5 correspond to the 7-AGNR whereas rectangles R1 and R4 correspond to the 6-AGNR. R3 (intensity multiplied by 10) most likely originates from 5-AGNR (see below).
The bulky feature around 1600\;cm$^{-1}$ comes mainly from a superposition of the the G line of the nanotubes and the AGNR. The circled signals 
originate from double-walled carbon nanotubes grown during transformation. They correspond to the weak Raman lines in Fig.\;\ref{fig:Temp}e,f assigned by the arrows. 
b, Experimental Raman excitation profiles 
derived for the Raman peaks that are indicated by the dashed white lines in panel a are presented as black squares together with the fitted Raman excitation profiles (in red) for the 6-AGNR (left) and 7-AGNR (right) (See also Supporting Information section (e)). The profiles correspond to the CH-ipb and D-line, respectively.
 }
\label{fig:Map}
\end{figure*}

{\bf Raman Excitation Profiles} Raman excitation profiles were measured for all Raman 
lines in the low and in the high frequency region for the excitation wavelength range from 400 - 800\;nm with 5\;nm steps. 
 Figure\;\ref{fig:Map}a shows overall results in the form of a two-dimensional Raman map for a sample transformed at a temperature of 850\;$^\circ$C. 
Raman map means Raman intensities are plotted on a color code versus Raman shift
and excitation energy.
The procedure to create the Raman map from the experiments and to extract the excitation profiles for the various modes by wavelength-dependent fitting of the experimental data, is described in detail in the Supporting Information Section (e).
%The features relevant for the Raman lines from the ribbons are collected into the white framed rectangles with 
%numbers R1 to R5. 
%The features outside the white rectangles are from the nanotubes. 
%The lower part of the map (R1 and R4) with features between 1,7 and 2,1 eV excitation 
The features between 1.7 and 2.1 eV excitation in the lower part of the map (R1 and R4)
correspond to the 6-AGNR. The upper part has 
the resonances of the 7-AGNR (R2 and R5). R3 is the resonance of a weak and as yet unknown new line to be discussed below. 
%To pin down the Raman lines in the map a simultaneous fit of all spectra at different laser excitation wavelengths was performed, using a sum of 
%Lorentzians with shared peak positions and line widths but varying Raman intensities as  
 
%More details of this simultaneous fit procedure are reported in the Supporting Information. 
%The so-obtained Raman frequencies and line widths are reported in Tab.1 and the Raman intensities for a select number of modes are depicted in Fig.4 (other modes are %shown in Supporting Information).
For a selected number of modes such as the CH-ipb and the D line, experimental 
resonance profiles are plotted in Fig.\;\ref{fig:Map}b. The profiles for the 
6-AGNR show only two vibronic peaks. 
%with some additional small peaks at higher energies.
%response in the energy range between 2,3 and 2,6 eV (not shown) for the D line and for the line at 1272\;cm$^{-1}$. 
%The latter could originate from defect states such as e.g. from end effects \cite{Cardoso18EPJB}.
In contrast, the resonances for 7-AGNR exhibit a highly 
structured profile which turned out to originate from vibronic sidebands as well as from higher exciton transitions. 
%At the low frequency edge of the resonance curves very weak structure are observed.
\par
The red lines in Fig.\;\ref{fig:Map}b are fits from the Kramers-Heisenberg-Dirac (KHD) theory for resonance Raman scattering \cite{Kiefer95Raman,Haroz11PRB} with wave functions in the adiabatic approximation.
%Application of this vibronic 
%theory is justified since ribbons are quasi-one dimensional systems with narrow energy bands. 
In this case the resonance scattering intensity $I_s$ is given by
\begin{equation}
I_s= CI_0\omega_s^4\sum_{\rho\sigma}|(\alpha_{\rho \sigma})_{fi}|^2 \qquad {\rm with} \\
\label{1}
\end{equation}
\begin{equation}
 (\alpha_{\rho \sigma})_{fi} =\sum_r{\frac{\langle f|\mu_\rho|r\rangle\langle r|\mu_\sigma|i\rangle}{\omega_{ri}-\omega_L-{\rm i}\gamma_r}} \nonumber
\end{equation}
where $I_0, (\alpha_{\rho \sigma})_{fi}, \mu_{\rho,\sigma}, \omega_{ri},$ and $\gamma_r$ are the intensity of the incident light, the transition polarizability, 
the dipole moment, the transition energy from state i to state r, and an electronic damping constant, respectively. 
%An additional non-resonant term was dropped in the above equation.    
Within the adiabatic approximation for the wave functions and a transition dipole moment independent of the vibrational normal coordinates 
(Albrecht A-term \cite{Tang70Raman}) this results eventually in the relation
\begin{equation}
(\alpha_{\rho \sigma})_{fi} =A= \sum_{e,\nu_e}{\frac{\mu^e_{0\rho}\mu^e_{0\sigma}}{\omega_{ri}-\omega_L-\rm{i}\gamma_r}}\langle \nu_f|\nu_e\rangle\langle\nu_e|\nu_i\rangle\; .
\label{2}
\end{equation}
$\mu^e_{0\rho,\sigma}$ are the $\rho$, $\sigma$ components of the pure electronic transition dipole moments which are assumed to be constant. $r$ stands as abbreviation for 
the transitions to states $e, \nu_e$.
The right part of the expression in Eq.\;2 depicts the vibronic matrix elements given by the Frank-Condon (FC) integrals. The 
vibronic quantum numbers $\nu_i$ were assumed zero except for i=0. This means 
%according to the low temperature in our experiments 
all electrons are in the vibronic ground state. $\nu_f$ was assumed 1 throughout meaning that only one vibron processes were considered. The experimental results for the 7-AGNR required 
electronic transitions up to $e=2$, i.e., in solid state terminology E$_{11}$ and E$_{22}$ were needed for the fit and thus experimentally determined. The most relevant parameters for the fit are the 
%exciton 
transition energies to the first and to the second excited state, 
 the Huang-Rhys factors, and the vibrational frequencies in these states.  
 Values for the transition energies are depicted in Tab.\;2. Excited state frequencies as they are listed
 in Tab.\;1 were found to be very close to the ground 
state frequencies. 
%They differ only by less than 5\% from the values for the ground state which justifies a straight forward application for the evaluation of the FC integrals. 
For the resonance of 
the CH-ipb mode in the case of 7-AGNR the line width in the excited state for the transition E$_{11}$ 
was considerably smaller than the line width in the ground state. Therefore the peak for the high energy resonance (outgoing resonance) 
is higher than the peak for the low energy resonance (incoming resonance). All parameters for the fitted resonances are summarized in the Supporting Information Section (f), Tab.\;S1. 
The observed transition energies were compared to values calculated by solving the Bethe-Salpeter equation within the frame of a 
quasiparticle self-consistent GW calculation. 
These calculations go beyond the energy dependent Placzek approximation as it was used to derive the Raman spectra described in Fig.\;2.
The calculation is described in detail in Methods and in Supporting Information Section (b). 
Table\;2 lists the observed transition energies as 
compared to our calculations and to reported values from optical reflection measurements 
\cite{Denk14NC}. 
%The very 
%good agreement between the observed energies and the values calculated for 7-AGNR particularly for the first transition (E$_{11}$) as well as the good 
%agreement with values reported 
%for such energies for 7-AGNR from STS measurements 
The comparison depicted in the table further evidences the successful growth of 7-AGNR and 6-AGNR inside SWCNT. 
\begin{table}
  \caption{Transition energies in eV as observed experimentally from resonance Raman analysis, compared to calculated values (in parentheses) and to references 
for AGNR. Experimental values were rounded to two digits behind the decimal point. Line 1 and line 3 were obtained for 7-AGNR from high frequency modes and from the RBLM, respectively. Column 5 and 6 are from references. 
%From the resonance of the RBLM even a transition energy beyond E$_{22}$ could be observed ($E_{33}$$^{'}$). 
The table also shows calculated transition energies for 5-AGNR, together with a value from a reference. 
	}
  	\tabcolsep0.5mm
  \begin{tabular}[htbp]{@{}llllll@{}}
    \hline
Ribbon& E$_{11}$ & E$_{22}$ &E$_{33}$ &E$_{11}$ & E$_{22}$ \\
      &Exp.  & Exp.    &            &Exp.&Exp.\\
			&(Calc.)&(Calc.)&&(Calc.)&(Calc.)\\
			&(eV)&(eV)&(eV)&(eV)&(eV)\\
    \hline
    7-AGNR &2.18 &2.45     &&2.1 Ref. \cite{Denk14NC}   &2.3 Ref. \cite{Denk14NC}   \\
		      &(2.37)&(2.75)  &(3.03)&(1.91)&(2.3)\\
		7RBLM&2.18&2.43&&&\\
		6-AGNR &1.83 &        &    &1.69 Ref.\cite{Merino-Diez17AN}& \\
		      &(1.82)& (3.02)&(3.24)&&\\
		5-AGNR& (0.84)&(2.19)&(3.02)&0.1 Ref.\cite{Kimouche15NC}&\\
    \hline
  \end{tabular}
  \label{tab:2}
\end{table}
%
%Resonances beyond E$_{22}$ could only be obtained for the RBLM. 
%They are assigned as $E_{33}$$^{'}$ as they exhibit noticeable 
%Since the corresponding energies exhibit a noticeable 
%difference to the calculated values for E$_{33}$ they might originate from highly disturbed higher transitions or from in-gap states \cite{Cardoso18EPJB}.
%The transition energies obtained from the fits and from the calculation for the first resonance peak for the 6-AGNR are listed 
%in Tab.\;2 line 4 and 5. Again the 
%agreement between experiment and calculation is excellent supporting the interpretation for the observed species to be 6-AGNR. 
\par
The two peaks in the resonance for the 6-AGNR represent the transitions to the first and to the second vibronic level in the excited state, 
or equivalently, the ingoing and outgoing resonance. Higher vibronic levels are neither observed for the 6-AGNR nor for the 7-AGNR. 
%not observed which is reasonable since ribbons are extended 
%objects and therefore are expected to have small Huang-Rhys factors. Likewise, the resonances for the 7-AGNR also exhibit only ingoing and 
%outgoing response. This is in difference to a response from molecules where higher vibronic sidebands are expected due to higher Huang-Rhys factors. 
\par
Calculated transition energies for 5-AGNR are included in the table since the values may be relevant for the resonance of the Raman line 
observed at 533\;cm$^{-1}$
 (R3 in Fig.\;3) as discussed below. The low value of 0.84\;eV calculated for the band gap is consistent with the 3p+2 family of these ribbons. 
\par
%%%%%%%%%%%%%%%%
%%%%%%%%%%%%%%%%%%%
%%%%%%%%%%%%%%%%%%%% RBLM
For the analysis of AGNR@SWCNT the RBLMs are particularly important since they exhibit a strong and characteristic response in a frequency region which is free from other Raman lines. 
%They are special in our case as they are expected to suffer strongly from the encapsulation. 
In the case of the 7-AGNR@SWCNT the response is unusual since it exhibits several components. 
Figure\;4a shows a zoomed-in Raman map of region R2. Several peaks can be observed in this region, which become also evident 
when plotting the Raman spectra for two distinct laser excitations (Fig.\;4b). The main peak (highest intensity) appears at a vibrational frequency of 
414\;cm$^{-1}$ for 569.5\;nm excitation with an exceptionally small line width of only 7\;cm$^{-1}$. In addition, a shoulder can be observed around 400\;cm$^{-1}$, which is considerably 
broader (13\;cm$^{-1}$) and exhibits dispersion, i.e. line position shifts with changing excitation energy. Such behavior is well known for the D-line in CNTs but also 
for conjugated polymers like poly-acetylene.\cite{Kuzmany82PRB} 
It is an indication for defective structures where vibrational frequencies and electronic transitions exhibit some correlation which eventually leads to photo-selective resonance scattering. 
In our case such structures may be represented by interfaces between extended or short ribbons of different structure and different topology. In such cases new electronic state inside the gap can be created.\cite{Cao17PRL}  
Also, modified edges resulting in sample inhomogeneity can lead to slightly modified 
RBLM-modes
%\cite{Overbeck19PSSB} 
and thus result in photo-selective resonance scattering. 
%{\it Such structural modifications are confirmed by the AC-HRTEM in Fig.1. However, a more detailed analysis for the %origin of the shoulder  is not possible at present. Also, the inherent dispersion of the corresponding Raman lines %make the determination of the resonance profile in this case difficult.}
%but suggesting that there are also defective ribbons in the tubes
In addition to the shoulder a very weak Raman signal is observed at 375\;cm$^{-1}$ for 506.1\;nm excitation and  
at shorter wavelength of 460\;nm a Raman line at 417\;cm$^{-1}$. At even shorter wavelengths of 405\;nm excitation a Raman line is observed at 533\;cm$^{-1}$. 
(The latter two are not shown in the figure.) The origin of these lines will be discussed below and in some 
more detail in Supporting Informations Section (e).  
\par
\begin{figure*}[tbp]
\includegraphics[width=0.85\linewidth]{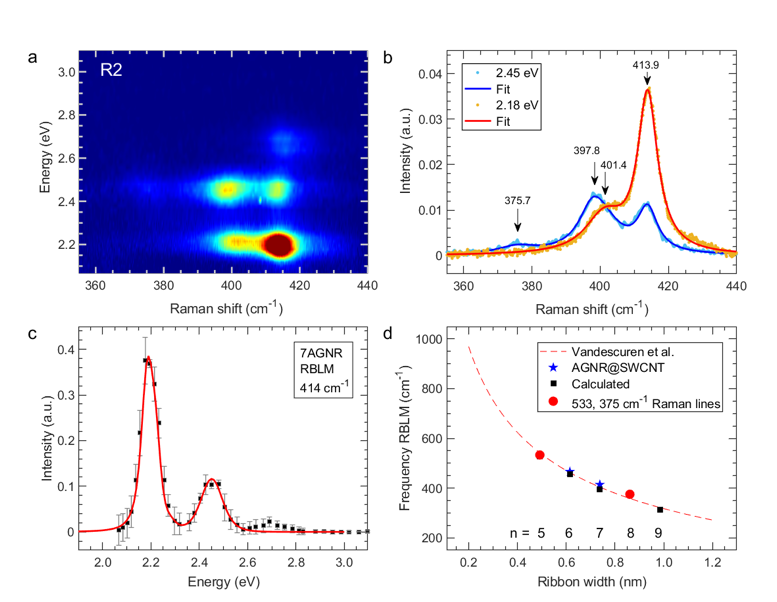}
\caption{Raman scattering for the RBLM of 7-AGNR. a, Blown up Raman map for the frequency region between 
350 and 440\;cm$^{-1}$;
b, Radial breathing like mode for 7-AGNR as excited with laser lines at 2,45 (blue) and 2,18\;eV (red). Full drawn lines are fits from the experimental analysis. c, Resonance profile for the main peak of the RBLM at 414\;cm$^{-1}$ (dots, with errorbars) together with the 
calculated values (red line); d, RBLMs frequencies versus ribbon width $w$. Blue stars are the experimental results from this work. 
Black squares are calculated values. The 
dashed line corresponds to Vanderscuren et al.\cite{Vandescuren08PRB} who found a relation of the form 
$\omega_{\rm RBLM} = a/\sqrt{w}-b$ with empirical parameters $a$ and $b$ for the dependence of the RBLM frequency on $w$, the ribbon width without hydrogen. The dashed line in the figure was evaluated for $a$ and $b$ equal to 527.4 and 210.2, respectively.   Red dots are for the additional 
Raman lines at 533 and 375\;cm$^{-1}$ if they are assigned to 5-AGNR and 8AGNR, respectively. The value for n = 9 is from a reference.\cite{Talirz16AM} 
}
\label{fig:RBLM}
\end{figure*}

   The strong line at 414\;cm$^{-1}$ is most appropriate to analyze resonance profiles. 
	%To extract this profile, region R2 was fitted with a sum of 5 Lorentzians 
%(see Supporting Information Section (e)). 
This profile is depicted in the Fig.\;\ref{fig:RBLM}c. It exhibits two strong resonances and a weak resonance at 2.7\;eV, which is not statistically significant and originates from a differentRaman frequency (see Supporting Information (e)). Due to the low mode energy for the RBLM(51 meV) the two resonances do not explicitly show vibronic splitting. Fitting the strong resonances with the 
Albrecht A term yields the transition energies listed in Tab.\;2. 
The resonances have very similar energies as they were observed for the high frequency 
modes of the 7-AGNR and are therefore assigned to E$_{11}$ and E$_{22}$. 
%The energy for the third peak at 2.65\;eV is considerably lower than the calculated value for E$_{33}$. 
%and much lower than reported values for E$_{33}$ in the literature \cite{Denk14NC}. 
%It is assigned to a higher transition involving electronic states inside the gap. Such states were recently observed %e.g. as originating for ribbon ends \cite{Cardoso18EPJB}. 
%The weak resonance at 2.7\;eV is not statistically significant (see Supporting Information (e)) and is therefore %not discussed in the following.
%
\par
RBLMs for AGNR are known to depend on the width $w$ of the ribbons. 
%In this sense they are the pendant to the radial breathing modes of carbon nanotubes. 
In several reports this dependence was observed to follow an $1/\sqrt{w}$ behavior \cite{Vandescuren08PRB}. 
%7sZhou. 
A more recent work showed that the RBLM was found to scale with 1/w.\cite{Liu20Carbon}  Within the small frequency range one cannot discriminate between the two scaling laws.
Figure\;\ref{fig:RBLM}d depicts this scaling law (Vandescuren diagram) as a dashed line. The black squares are the frequencies calculated here for 5-AGNR, 6-AGNR and 7-AGNR. 
The experimentally observed frequencies for 6-AGNR and 7-AGNR are indicated as blue stars. The good fit of these values into the Vandescuren 
diagram is another proof for the observation of such ribbons in our samples. 
Interestingly the so far unassigned lines at 533\;cm$^{-1}$ and 375\;cm$^{-1}$ fit also very well to the diagram if they are assigned to RBLMs for
 5-AGNR and 8-AGNR. Recently the RBLM for 5-AGNR was observed at 533\;cm$^{-1}$.\cite{Chen17JACS,Overbeck19PSSB} However, since both ribbons are narrow gap species of the type 3p+2, 
E$_{22}$ resonances or even higher transitions must be involved for their observation. Interestingly for 8-AGNR very recently an effective gap of 2.3 eV was 
reported from STS measurements.\cite{Sun19Small} This is reasonably close to the 2.45 eV laser used to detect the Raman line at 375\;cm$^{-1}$. The Raman line at 533\;cm$^{-1}$ was observed in our case for excitation with a considerable higher energy than the value calculated for the $E_{22}$ transition.
Shortening of the ribbons with respect to infinite long species could be the reason for the enhanced transition energies\cite{Kimouche15NC,Talirz19CPC} as they are observed here. 
 
\section {Discussion and Conclusion}
{\bf Discussion}

When using polyaromatic hydrocarbons as precursor molecules (such as coronene\cite{Chernov13AN,Anoshkin14CPC}), the growth of GNRs inside CNTs can be explained by an oligomerization reaction. In our case, however, and similarly as previously reported for tetrathiofulvalene molecules inside CNTs,\cite{Chamberlain12AN} the encapsulated ferrocene molecules first need to be decomposed by the heat treatment and subsequently need to form the nanoribbons. This process is assumed to benefit from the graphene-like walls of the nanotubes, but the exact process is yet unknown. Until now, only graphene nanoribbons with armchair structures were observed, which is probably due to the higher stability of these ribbons. The dominance of the growth of 6AGNR and 7AGNR is due to the mean diameter of the carbon nanotubes used. Ribbons with a higher width would not fit into the tubes without extra stress and too narrow ribbons would would be energetically less favorable. 
\par
The ribbons observed in our study are definitely hydrogen passivated. This can be concluded from the strong CH-ipb mode observed in the Raman spectra. The strong response of the hydrogen vibration is also reflected in the ab initio calculations. Passivation by more extended hydrocarbon groups is not likely due to the lack of space inside the tubes.
\par
The concentration of the ribbons inside the tubes can be estimated from the available carbon resources and thus the degree of filling of the tubes with ferrocene. A (17,0) tube can be filled with approximately 4 molecules of ferrocene per nm,\cite{Plank10AN}  which means 40 C/nm can be provided. A 7-AGNR needs 33.3 C/nm, hence providing sufficient carbon atoms to form the graphene nanoribbons and thus a high concentration of ribbons can be expected. We indeed observe very high Raman intensities from the AGNRs compared to the Raman signatures of the surrounding CNTs, however, the Raman cross-section of encapsulated AGNRs with respect those of CNTs is not yet known. Hence, one cannot use the Raman intensities to estimate the synthesis yield.

\par
 The TEM analysis gives clear evidence for flat objects with ribbon structure inside the tubes. The deviation from a perfect graphenic ribbon structure observed in the figure is a consequence of the electron beam irradiation rather than being intrinsic to the ribbon growth.
The identification of the ribbons  comes from wavelength-dependent Raman scattering and comparison to high level quantum-chemical calculations. 
The very narrow Raman lines as depicted in Tab.\;1 and Fig. 2 are evidence for clean and highly unperturbed ribbon material. 
 For the 7-AGNR the experimental and theoretical Raman spectra exhibit unprecedented agreement with respect to 
frequency of the modes and the relative Raman intensities. 
\par

The very sharp and individual peaks of the resonance profiles are evidence for a well-defined edge structure, which does not allow for defect induced modulation and consequently broadening or splitting of the optical transitions.
%{\it The analyses of the resonance excitation profiles were performed with high effort. (See also Supporting Information Section (e)).  They allow for a look at the ribbons buried in the tubes and provide information on their electronic structure. In contrast to the optically opening of the nanotubes by functionalization \cite{Lim15AN} this method is completely nondestructive.}
%{\it Extended sequences of geometrically and topologically different ribbons can be connected  in the same tube.}
\par
{\bf Conclusion}

The results presented above demonstrate that graphene nanoribbons can be grown from thermal decomposition 
of ferrocene encapsulated in carbon nanotubes. Aberration corrected high resolution transmission electron microscopy provides 
evidence for the ribbons inside the tubes. Raman scattering combined with first principle calculations and resonance excitation analysis 
is an excellent tool to identify the structure of GNR even if the ribbons are encapsulated in SWCNTs. 
Due to its nondestructive nature and frequency selective character the Raman method is superior to any other optical method such as optical absorption or luminescence spectroscopy.
In addition the method allows determining the electronic structures of the ribbons, even beyond the HOMO-LUMO gap.
The results provide a challenge that ribbons with larger or smaller width than those reported here, can be grown in larger or smaller SWCNTs, 
respectively, and can be detected with Raman scattering. They thus open a new field in subnanometer graphene ribbon research.
%For example, according to Tab.\;2  n=5 AGNR can be observed even with red lasers 
%if the resonance to the LUMO+1 state (E$_{22}$ transition) is used. 
%In addition, the results from Fig.\;\ref{fig:RBLM} 

\section{Methods}
{\bf Synthesis of GNRs@SWCNT} SWCNTs as grown by the eDIPS technique \cite{Saito08JNN} with mean diameter around 1.3 nm were used as a 
starting material. All tubes were purified by first etching in air and subsequent treating with HCl as described previously 
\cite{Kuzmany14PSSB} to remove amorphous carbon and catalytic particles from tube growth. SWCNT bucky paper was obtained by washing 
and filtering the tubes with distilled water and ethanol. 
\par
For the filling process the tubes were first opened by etching in air at 420\;$^\circ$C and then exposed to ferrocene at 400\;$^\circ$C for two 
days in previously evacuated quartz tubes. The transformation of the FeCp$_2$ molecules to the AGNRs inside the tubes was performed 
by vacuum annealing for several days and at various temperatures as described in the main text. This process resulted in the appearance of new 
lines in the Raman spectra from some so far unknown objects. Control experiments were performed without opening the tubes but otherwise 
treating the material identically. In this case almost no new Raman lines were observed as depicted explicitly in the Supporting Information Section (c). This can be considered as evidence for the growth of the objects inside the tubes. 

{\bf Analysis by Transmission Electron Microscopy} To obtain information on the grown objects TEM investigations were performed with a Thermo Fisher (formerly FEI) TITAN G2 80-300 (Fig.\;1a-e) 
at 80 kV. Figure 1f, g and i are acquired by the specific Thermo Fisher (formerly FEI) SALVE
(subangstrom low voltage electron microscopy)
transmission electron microscope fitted with a CEOS CETCOR  spherical aberration corrector (axial 5th order, off-axial 3rd order), a chromatic 
aberration corrector,  and a Thermo Fisher Ceta 4K CMOS camera. 
%The microscope was operated at an electron acceleration voltage of 80 kV 
%with a point resolution smaller than 0.08 nm. 
The exposure time of the images is 1.0 s and the dose rate is 6.85*10$^6$ electrons s$^{-1}$ nm$^{-2}$.

{\bf Theoretical evaluation vibrational frequencies and mode-specific Raman intensities}
Vibrational frequencies of the ribbons were calculated at the $\Gamma$ point using the Vienna Ab initio Simulation Package (VASP).\cite{Kresse96PRB} Raman intensities $I_s$ were evaluated using the frequency dependent Placzek approximation. 
In this case the intensity of the Raman lines is proportional to the square of the derivative of the frequency dependent polarizability with respect to the phonon normal mode:
\begin{equation}
 I_s(\omega_s,\omega_L)=\frac{\omega_s^4}{\omega_L}\sum_{\rho,\sigma}\left|\frac{\partial\alpha_{\rho,\sigma}(\omega_L)}{\partial Q_{ph}}\right|^2
\Gamma(\omega-\omega_{ph})(n(\omega_{ph})+1)
\label{eq:3}
\end{equation}
where $\alpha_{\rho,\sigma} (\omega_L)$ is the dynamic (frequency dependent) polarizability tensor  evaluated from first principle calculations at the laser energy $\omega_L$. Calculation of the dynamic polarizability was done within the linear response theory \cite{Gajdos06PRB} with wave functions and structural parameters used during the frequency calculation and geometrical optimization.  $Q_{ph}$ and $\omega_{ph}$ are the phonon normal modes and 
frequencies, respectively, and $\omega_s$  is the frequency of the scattered light. $\omega$ is the difference between $\omega_L$ and $\omega_s$.
 $n(\omega_{ph})$ is the Bose-Einstein distribution at room temperature and $\Gamma(x)$ is a 
normalized Lorentzian function with full width at half maximum of 10\;cm$^{-1}$. Numeric derivatives of the polarizability tensor were 
calculated using symmetric derivatives by manually shifting the atoms according to normal modes in both positive and negative directions. 
From this the following expression can be obtained for the derivative of the polarizability:
\begin{equation}
\frac{\partial\alpha_{\rho,\sigma}(\omega_L)}{\partial Q_{ph}}=\sum_{r,s}{\frac{\langle f|H_{e-p}|s\rangle\langle s|H_{e-ph}|r\rangle\langle r|H_{e-p}|i\rangle}{(\omega_L-\omega_{r}-{\rm i}\gamma_{e-p})(\omega_r-\omega_s-{\rm i}\gamma_{e-ph})}}
\label{eq:4}
\end{equation}
where r,s are intermediate virtual electronic states with energy $\omega_{r,s}$ obtained from first principle calculations, $H_{e-ph}$ (derivative of the electron-ion potential with respect to the normal modes) and $H_{e-p}$ are the Hamiltonians for the electron-phonon and electron-photon  coupling, respectively and $\gamma_{e-ph}$ and $\gamma_{e-p}$ are the corresponding damping constants (life times). 
Details of the derivation can be found in ref. \cite{Kukucska19arX}. 
%This is our implementation for VASP, 
The calculation was successfully used previously for the analysis of doping and strain induced changes of the Raman spectra and gap structure of MoS$_2$ \cite{Peto192DM} and silicene.\cite{Kukucska18PRB}
%Similar approaches are reported recently by other authors [other1,other2,other3].
\par
Optical excitation energies were calculated by solving the Bethe-Salpeter equation within the frame of a quasiparticle self-consistent 
GW calculation as built in the QUESTAAL code \cite{Kotani07PRB} with ladder diagram corrections. 
\cite{Cunningham18PRM} Excitation energies were determined by taking the energy values at the maximum of the peaks 
present in the imaginary part of the macroscopic dielectric function. Explicitly, calculations were performed for the low gap 3p+2 
ribbon 5-AGNR, for the moderate gap 3p ribbon 6-AGNR, and for the large gap 3p+1 ribbon 7-AGNR. More details on the calculation are in supporting Information Section (b).
\par
{\bf Raman scattering} Raman spectra were excited at room temperature and at ambient conditions with various lasers in the visible spectral range. Spectra were recorded with 
a Dilor xy800 and with a Labram HR800 microscope. In Figure 2 spectra  with a spectral resolution of less than 2 wavenumbers. Thus the observed linewidths are intrinsic to the ribbon material. Raman intensities were normalized to the 2D line of the carbon nanotubes, since the 2D-line is known to be highly independent from defects. In the case of resonance analysis normalization was performed to the fundamental Raman line of Si. 
\par
{\bf Evaluation of Excitation Profile} Resonance Raman spectra were evaluated for wavelength dependent excitation in the range 
between 400 and 800 nm with 5 nm step-size. 
To cover the requested energy range for excitation the following three laser systems were used.
400-526 nm: A frequency-doubled Ti:Sa laser (Msquared SolsTis external cavity frequency doubled ECD-X module)
 which was pumped by an 18 W Sprout-G diode pumped solid state laser (532 nm).
534-605nm and 610-690 nm: A dye laser (spectra Physics model 375) pumped by an Ar+ ion laser (Spectra physics 2020) equipped with either 
Rhodamine 110 or DCM laser dyes. 690-800 nm:  A tunable Ti:Sa laser (Spectra Physics 3900S).
The Raman spectra were recorded with a high resolution triple grating Dilor XY800 spectrometer and a liquid nitrogen cooled CCD detector. 
More details about the recording and evaluation of the Raman spectra can be obtained from Supporting Information Section (e).


\section{Acknowledgement}
Work supported by the NSFC (51902353), the FWF project P21333-N20, and the NKFIH, Grant No. K-115608. 
L.S. acknowledges the financial support from the Natural Science Foundation of Guangdong Province (Grant No. 2019A1515011227) and the Sun Yat-Sen University (Grant No. 29000-18841218, 29000-31610028)
J.K., J.K. and G.K. further acknowledge the [NIIF] for awarding access to 
resource based in Hungary at Debrecen, they further acknowledge support by the National Research Development and Innovation Office of Hungary within the Quantum Technology National Excellence Program  (Project No. 2017-1.2.1-NKP-2017-00001), and the ELTE Excellence Program (1783-3/2018/FEKUTSTRAT) 
supported by the Hungarian Ministry of Human Capacities.  
K.C. acknowledges the China Scholarship Council (CSC) for financial support. U.K. acknowledges the support of the Graphene Flagship and DFG SPP Graphene as well as the DFG and the Ministry of Science, Research and the Arts (MWK) of Baden-Wuerttemberg within the frame of the SALVE project.
M.M., S.C., and W.W. acknowledge funding from the Fund for Scientific Research Flanders (FWO projects No. G040011N, G02112N, G035918N, G036618N and the EOS-charming 
project G0F6218N [EOS-ID 30467715] ). M.M. acknowledges funding of a DOCPRO4 PhD scholarship from the UAntwerp research fund (BOF) and S.C. also acknowledges 
funding from the European Research Council Starting Grant No. 679841.

\section{Author Contributions}
H.K. and L.S. contributed equally to this work. H.K. and L.S. designed and supervised the experiments. L.S. prepared the samples and  did the characterization with spontaneous Raman scattering in Vienna with lasers at wavelength of 633 and 568 nm. S.C., M.M., and W.W. performed the wavelength-dependent resonance Raman scattering experiments and their analysis. H.K. analyzed the resonance profiles. K.C. and U.K. performed HRTEM characterization and simulations. J.K., J.K., and G.K. performed the first principles DFT calculations. T.S. provided the SWCNTs. T.P. provided the laboratory facilities and the Raman setup in Vienna. All authors discussed the results and commented on the manuscript at all stages. \vspace{1cm}

\section{Competing Financial Interests}
The authors
declare that they have no competing financial interests associated
to the publication of this manuscript.

\indent {\large \bf Corresponding Author}
hans.kuzmany@univie.ac.at, shilei26@mail.sysu.edu.cn

\indent {\large \bf References}
%\bibliography{GNRs}

\end{document}